# A theory of necking in semi-crystalline polymers

A. I. Leonov

*Department of Polymer Engineering, The University of Akron,*

*Akron, Ohio 44325-0301, USA.*

**Abstract**

Necking or cold drawing is a smoothed jump in cross-sectional area of long and thin bars (filaments or films) propagating with a constant speed. The necks in polymers, first observed about seventy years ago, are now commonly used in modern processing of polymer films and fibers. Yet till recently there was a lack in fundamental understanding of necking mechanism(s). For semi-crystalline polymers with co-existing amorphous and crystalline phases, recent experiments revealed that such a mechanism is related to unfolding crystalline blocks. Using this idea, this paper develops a theoretical model and includes it in a general continuum framework. Additionally, the paper explains the "forced" (reversible) elasticity observed in slowly propagating polymeric necks, and also briefly analyses the viscoelastic effects and dissipative heat generation when polymer necks propagate fast enough.

## 1. Introduction

The necking phenomena usually occur when a homogeneous solid polymeric bar (film or filament), with a non-monotonous dependence of axial force $S$ on stretching ratio $\lambda$, is stretched uniaxially in the region of $S(\lambda)$ non-monotony. In this case the polymer bar is not deformed homogeneously. Instead, two almost uniform sections occur in the sample: one being nearly equal to its initial thickness and another being considerably thinner in the cross-sectional dimensions. These sections are jointed by a relatively short transition (necking) zone that propagates with a constant speed along the bar as a stepwise wave in the direction of the bar's thick end (Fig.1).

The following fundamental questions should be answered by any theory attempting to quantitatively explain the necking phenomena in physical terms.

(i) What is the stabilization mechanism that keeps constant the speed of neck propagation?

(ii) What is the physical reason for the $S - \lambda$ non-monotony?

(iii) What is the physical mechanism of the forced elasticity?

(iv) How the viscoelasticity and dissipative heat generation do affect polymer necking?

Carothers and Hill (1932) were first who discovered the necking phenomena. They observed the necks propagating in filaments of semi-crystalline polyester near the room temperature. Whitney and Andwrews (1967), Crissman and Zapas (1974), Zapas and Crissman (1974) found later that under certain conditions necking could also occur in glassy polymers. Kozlow, Kabanov and Frolova (1959) studied the temperature effect on necking for the semi-crystalline polymers and Bartenev (1964) did that for and glassy polymers. Some additional data were reviewed in monographs by Kargin & Slonimsky (.1967), Gul' and Kulizniov (1972) and Tager (1978). Many of these publications showed that the high stretching ratios achieved usually in cold drawing were in many cases almost reversible in nature, and some specimens after heating to a solid-state temperature completely recovered their initial form and dimensions (e.g. see Lazurkin and Fogel'son (1951) Lazurkin (1958), and more recently Gent and Jeong (1986), Gent and Madan (1989)). Lazurkin (1951, 1958) called this effect was *forced elasticity*.

Orowan (1949) and Nadai (1950) proposed a formal, mechanistic explanation of cold drawing in terms of the $S(\lambda)$ dependence obtained for homogeneous elongation. A detailed discussion of this viewpoint can be found in the text by Ward (1982), Ch. 11.

Barenblatt (1964) was the first who suggested a theory of polymer necking. In the spirit of the flame propagation theory, he proposed a quasi-1D approach taking into consideration a special "stress-diffusion" phenomenon that was theoretically necessary in his approach for stabilizing the neck propagation. Barenblatt treated the necking as a boundary-value problem for a nonlinear second order ODE whose eigenvalue was the speed of the neck propagation. However, his principal assumption of the stress-diffusion did find no experimental basis.

A distinct mechanism explaining the $S(\lambda)$ non-monotony due to a local adiabatic heating was qualitatively proposed by Muller (1949) and later by Marshall and Thompson (1954). Barenblatt, Entov and Segalov (1968) developed a related 1D theory in which the mathematical problem was treated similarly to Bareblatt (1964), but with the stabilizing, longitudinal heat conductive term. Further studies of cold drawing by Brauer and Muller (1954), Vincent (1960) and Allison and Ward (1967) demonstrated, however, that at low rates of extension common for technological applications, the adiabatic increase in temperature is too small to explain the necking. Lazurkin (1958) was the first who noticed this fact when performing the experiments at very low stretching rates.

To describe the cold drawing in terms of nonlinear elasticity, Antman (1973,1974) conceptualized the importance of non-uniform deformations in necking. In spirit of the approximate theory for long elastic bar, he analyzed a particular 3D field of strain, neglecting the tangential stresses. Using an averaging procedure and introducing strain energy functional, Antman reduced the problem to finding minimizers of the functional and studying the stability and bifurcation conditions. Later this program was fulfilled by Owen (1987) in a very rigorous way.

Ericksen (1975) discussed the non-monotony in the $S(\lambda)$ dependence and demonstrated a remarkable similarity between the necking and phase transition in the van der Waals' gas. Using this similarity, he proved that the constant equilibrium ("Maxwell") force could be found from energetically equal presenting both the oriented and disoriented "phases".

Coleman (1981,1985) and later Coleman and Newman (1988, 1990,1992) incorporated in the theory a 1D inhomogeneity in the necking region, which occurs in the cold drawing. Similarly to Antman (1973,1974) and Owen (1987), they considered a particular case of 3D finite elastic deformations within an approximate theory valid for long slender bars. It allowed them to include the terms proportional to the longitudinal gradient of stretching ratio $\lambda_z$ into the formulation of the elastic Gibbs' free energy functional Then the equilibrium equations obtained using the standard variational procedure, naturally involved the stabilizing term $\lambda_{zz}$.

The Erickson (1975) theory was extended by Bernstein and Zapas (1981) for the case of viscoelastic solid bars with the use of the BKZ viscoelastic constitutive relations. It was shown that postulated multiple valued dependence of viscoelastic force could have a tendency to destabilize the neck propagation. Later Coleman and Newman (1992a) proposed a general inhomogeneous 1D theory of cold drawing in viscoelastic solid materials.

Several problems of cold drawing were solved numerically. Needleman (1972) and also Burke and Nix (1979) analyzed the hypothetical plasticity effects in necking. The numerical solution by Silling (1988) of 2D elastic necking problem with assumed $S(\lambda)$ non-monotony demonstrated the closeness between the Maxwell and calculated actual necking forces.

It should be mentioned that except for the author paper (Leonov, 1990) that analyzed the surface energy effects as the main cause for necking in nano-size filaments, all the abovementioned theoretical and numerical works have never attempted to describe the physics of the $S(\lambda)$ non-monotony, as well as the forced elasticity effects in polymer necking.

On the other hand, qualitative structural models of necking have been intensively discussed in experimental papers. The most popular such a model proposed by Peterlin (1966) considered the folded chain blocks in necking of semi-crystalline polymers as tilted, sheared, broken off the lamellae and become incorporated in the (amorphous) microfibrils. Gent and co-workers (1986,1989) (see also earlier discussions by Statton (1967) and also Juska and Harrison (1982,1982a)) have recently proposed another model for semi-crystalline (SC) polymers. They related necking to the mechanism of unfolding chains in crystalline blocks and transferring them into amorphous phase with consequent orientation. Thus this model explains the necking by mechanical melting of the folded chain blocks. It also explains the puzzling fact that the higher the degree of crystallinity the higher is the necking final stretching ratio. Although some easy consequences of this modeling were also exploited by Gent and co-workers (1986,1989), the model has not been developed theoretically. Even for easiest situation of quasi-elastic necking regime, there was no attempt to derive the non-monotonous constitutive dependence $S(\lambda)$.

It should also be noted that to the author's knowledge, no physical model, even in qualitative sense, has been proposed for necking of glassy polymers, except vague mentioning the shear bands caused, perhaps, by crazing (e.g. see Gent and co-workers (1986,1989)).

From the above literature survey is clear that the only item (i) from the list of fundamental problems for polymer necking has been resolved. Therefore this paper is focused on the resolving remaining problems (ii)-(iv) from the list. The main objective of this paper is description of the $S(\lambda)$ non-monotony. This is achieved with theoretical developing of the qualitative model proposed by Gent and co-workers (1986,1989), and incorporating the result into the general mechanical frame established by Ericksen (1975) Coleman (1981,1985) and Coleman and Newman (1988,1990,1992) for quasi-elastic description of necking. Another objective is including viscoelastic and heat effects into a simple theoretical scheme.

The structure of the present paper is as follows. The second Section develops the kinetic model of unfolding the crystalline blocks in SC polymers. The third Section uses this kinetics in formulating the key function $S(\lambda)$ and discusses the Maxwell stress in the model. The fourth Section analyzes the slow (quasi-static) neck propagation. The fifth Section briefly considers the forced elastic irreversible effects. Some model restrictions and other effects in cold drawing of semi-crystalline polymers that are not described by the proposed theory are discussed at the end of the paper.

## 2. Kinetics of unfolding crystalline blocks in simple elongation of SC polymers

We employ here a simplified model of Gent and co-workers (1986,1989) for initial structure of SC polymers as a set of randomly oriented plane crystalline blocks, two ends of which are connected to the amorphous phase (Fig.2). The random orientation of crystalline blocks is caused by preliminary crystallization of samples under quiescent conditions. Thus the semi-crystalline polymer is simplistically modeled as consisted of amorphous (glassy) parts, "cross-linked" by the crystalline blocks.

Consider now a long axisymmetric bar (a film or rod) of a SC polymer, uniaxially stretched in the longitudinal direction. Let $c_0$ be the degree of crystallinity in a virgin sample. Let a non-deformed crystalline block be consisted of an averaged value $n_0$ monomer linear aggregates (or the monomer units themselves) united in a folded part of macromolecule as shown in Fig.2. The secondary interaction between the monomer units in the crystalline block is described by the periodically located potential wells of the width $h$ and depth $D_0$, with the periodicity $H$ (Fig.3). Here $D_0$ is the energy of dissociation of secondary bonds between monomers in the crystalline block, or the specific heat of fusion. Since the monomer units in the crystalline blocks are identical, the "chemical" parameters $h, H$ and $D_0$ are assumed not to fluctuate. If the total longitudinal force $S$ applied to the polymer bar exceeds a certain critical value $S_*$ (which is still less than the averaged "strength" of chemical bonds along the polymer chain), the crystalline blocks began unfolded and supply the unfolded parts of macromolecules to the amorphous phase. As soon as a part of the crystalline block is unfolded, it should travel a variable distance $l(t)$ of the order of Kuhn segment $l_k$ (the segment AB in Fig.3), to be incorporated in the local oriented environment in amorphous phase. The unfolding process is assumed to be slow enough to be viewed as non-dissipative. It means in particular, that the total strain of the bar during and after necking is reversible.

Consider now approximate kinetics of unfolding crystalline blocks, when $S > S_*$. Let $n(t)$ be the averaged value of the unfolded aggregates, so that $n(0) = 0$. Let $f(t)$ be the local fluctuating force acting on the unfolding chain, and $l(t)$ the fluctuating distance from the current unfolding site to a moving point in amorphous phase, close to the crystalline block (Fig.3). The energy balance for unfolding process is:

$$D_0 dn/dt = 2 < f \cdot dl/dt > . \tag{1}$$

The left-hand side of Eq.(1) presents the rate of dissociation of unfolding crystal and the right-hand side, the time-space average of the rate of work produced by two unfolding forces $f(t)$ acting at the ending chains of crystalline block. Because of very high, glass-type viscosity in the amorphous phase near the block, the possible rotation of the crystalline block is neglected. In order to calculate the right-hand side in Eq.(1) we use the physically motivated scaling approximations:

$$<f \cdot dl/dt> = <fh \cdot h^{-1} \cdot dl/dt> \approx D_0(l_k/h)<l^{-1}dl/dt>. \qquad (2)$$

Here the scaling relations, $fh \sim D_0$ and $l(t) \sim l_k$ have been used. We now can evaluate the averaged term in the right-hand side of Eq.(2) as follows:

$$<l^{-1}dl/dt> \approx c_0(n_0 - n)<k_z>_0 \lambda^{-1}d\lambda/dt. \qquad (3)$$

Here $\lambda$ and $\lambda^{-1}d\lambda/dt \equiv \dot{\varepsilon}$ are respectively the macroscopic stretching ratio and stretching rate, and $<k_z>_0$ is the longitudinal, $z$-axis, component of the unit orientation vector $\underline{k}$, characterizing the initially oriented crystalline blocks. Assuming a uniform distribution of initially oriented crystalline blocks yields: $<k_z>_0 = 1/2$. Substituting this value and also Eqs.(2) and (3) into Eq.(1) divided by the value $n_0$, results in the kinetic equation for the unfolding process:

$$-d\alpha/dt = m\alpha\lambda^{-1}d\lambda/dt; \quad \alpha \equiv 1 - n/n_0; \quad m = c_0 l_k/h. \qquad (4)$$

Here $\alpha(t)$ is the averaged portion of existing (not destroyed) crystal in the blocks, so $c(t) = c_0\alpha(t)$ is the actual degree of crystallinity at time $t$. Integrating Eq.(4) with the natural initial conditions, $\alpha(0) = 1$ and $\lambda(0) = 1$, yields the remarkable simple result:

$$\alpha(t) = \lambda^{-m}(t). \qquad (5)$$

## 3. Modeling finite elasticity of SC polymers

This Section models macroscopic finite 3D and 1D elongation elastic deformations of SC solid polymers. For the sake of simplicity, we further assume these polymers to be incompressible. To develop a continuum approach, which captures essential features of two-phase, crystalline/amorphous, SC polymers, we model initially the mechanical behavior of each phase, which is completely described by the respective strain energy function $F$ (the Helmholtz' local thermodynamic potential per mass unit) for 3D deformations. As assumed, the initial crystallites in samples of SC polymers are chaotically oriented. Therefore the finite elasticity in these materials could be considered as inherently isotropic.

The general 3D formulae of incompressible finite isotropic elasticity for the elastic potential $W$ $(= \rho F)$ and stress tensor $\underline{\underline{\sigma}}$ in the Eulerian, Cartesian presentations are of the form (e.g. see Truesdell and Noll (1992))

$$W = W(T, I_1, I_2); \quad I_1 = tr\underline{\underline{B}}, \quad I_2 = tr\underline{\underline{B}}^{-1}, \quad I_3 = \det \underline{\underline{B}} = 1; \quad \underline{\underline{\sigma}} = -p\underline{\underline{\delta}} + 2W_1 \underline{\underline{B}} - 2W_2 \underline{\underline{B}}^{-1}. \quad (6)$$

Here $T$ is the temperature, $\underline{\underline{B}}$ is the Finger strain tensor, $\underline{\underline{B}}^{-1}$ is the Cauchy-Green strain tensor, $I_k$ are the strain invariants, $p$ is the pressure, $\underline{\underline{\delta}}$ is the unit tensor, and $W_k = \partial W / \partial I_k$. The formula for stress in (6) is presented in the Finger form. In the case of simple elongation with the stretching ratio $\lambda$, the general formulae due to (6) for strain invariants, the actual elongation stress $\sigma$ and engineering stress $S$ ($= \sigma / \lambda = \partial W / \partial \lambda$) defined as elongation force per initial cross-section, are:

$$I_1 = \lambda^2 + 2\lambda^{-1}, \quad I_2 = 2\lambda + \lambda^{-2}, \quad \sigma = (W_1 + W_2 / \lambda)(\lambda^2 - \lambda^{-1}), \quad S = (W_1 + W_2 / \lambda)(\lambda - \lambda^{-2}). \quad (7)$$

The strain energy function $W_c$ for the crystalline phase is modeled here similarly to finite deformations of hard polycrystalline materials, such as metals, rocks etc. Macroscopic behavior of these materials can be considered as intrinsically isotropic, because of random orientation of their crystals. Thus the suitable dependence of the function $W_c$ can be searched from the class: $W_c = W_c(T, I_2)$. We will further use the simple 3D dependencies for the strain energy function $W_c$, and corresponding extra stress,

$$W_c = G_c(I_2 - 3)/2, \quad \underline{\underline{\sigma}}_c = -G_c \underline{\underline{B}}^{-1}. \quad (8)$$

Here $G_c$ is the Hookean modulus for the crystalline phase. Due to Eqs.(7,8), the formulae for strain energy function $W_c$ and force $S_c$ in simple elongation are of the form:

$$W_c = 1/2 G_c (2\lambda + \lambda^{-2} - 3), \quad S = G_c (1 - \lambda^{-3}). \quad (9)$$

Eq.(9) shows that with growing $\lambda$, the force $S_c(\lambda)$ rapidly reaches the upper constant value $G_k$. That was the reason that the potential function (8) has been chosen.

We now model the strain energy function $W_a$ and corresponding extra stress $\underline{\underline{\sigma}}_a$ for the amorphous phase similarly to that for finite deformations of cross-linked elastomers, taking into account the finite extensibility of macromolecular chains:

$$W_a = 1/2 G_a (J-3) \ln\left(\frac{J-3}{J-I_1}\right), \quad \underline{\underline{\sigma}}_a = G_a \frac{J-3}{J-I_1} \underline{\underline{B}}, \quad G_a = \rho RT / M_*. \tag{10}$$

Here $J$ ($\approx 100$) is a value of $I_1$ corresponding to the fully extended polymer chain, and in the formula for the Hookean modulus $G_a$, $\rho$ is the density, $R$ is the gas constant and $M_*$ is averaged molecular weight of parts of macromolecules between cross-links. The idea of the specific expression (10) for $W_a$ suggested long ago by Warner (1972), was used later for viscoelastic liquids in the text by Larson (1988), and recently proposed again by Gent (1996) for cross-linked elastomers. Due to Eqs.(7,10), the formulae for strain energy function $W_a$ and force $S_a$ in simple elongation are of the form:

$$W_a = 1/2 G_a (J-3) \ln\left(\frac{J-3}{J-\lambda^2-2\lambda^{-1}}\right), \quad S_a = G_a \frac{J-3}{J-I_1}(\lambda - \lambda^{-2}). \tag{11}$$

The strain energy function for 3D deformations of SC solid polymers is now proposed in the form:

$$W \approx c_0 \alpha(t) W_c(\underline{\underline{B}}) + [1 - c_0 \alpha(t)] W_a^*(\underline{\underline{B}}). \tag{12}$$

Here the lower indices "c" and "a" stand for crystalline and amorphous phases, $\underline{\underline{B}}$ is the total Finger strain in the SC polymer, and the symbols $c_0$ and $\alpha$ have been explained in the previous Section. The contribution of "amorphous" strain energy function $W_a^*$ in the total one $W$ is different from $W_a$ defined in Eq.(11). It reflects the fact that only a fraction of macromolecules with the average molecular weight, $M_a^* = M^*[1 - c_0 \alpha(t)]$, is involved in the amorphous phase. Therefore using formulae (11) in Eq.(12) with changing $G_a$ for $G_a^* \approx \rho RT / M_a^*$ yields:

$$W \approx c_0 \alpha(t) W_c(\underline{\underline{B}}) + W_a(\underline{\underline{B}}). \tag{13}$$

Here the term $W_a$ is the same as in Eq.(11), with the modulus $G_a$ presented in Eq.(10). Substituting now Eqs.(6, 9, 11) into Eq.(13), results in the formulae:

$$W = 1/2 G_c c_0 (2\lambda + \lambda^{-2} - 3)/\lambda^m + 1/2 G_a (J-3) \ln\left(\frac{J-3}{J-\lambda^2 - 2\lambda^{-1}}\right), \tag{14}$$

$$S(\lambda) = \partial W / \partial \lambda \big|_{\alpha=const} = G_c c_0 (1-\lambda^{-3})/\lambda^m + G_a \frac{J-3}{J-I_1}(\lambda - \lambda^{-2}). \tag{15}$$

Eqs.(14,15) represent the model of equilibrium (elastic) simple elongation for a SC polymer. It should be mentioned that these equations are invalid when the initial degree of crystallinity $c_0$ in SC polymers is either close to the unity or zero (less than gelation point for the cross-links), because in these cases the formulae of rubber elasticity (10) are not applicable. The plot $S(\lambda)$ according to Eq.(15) is sketched in Fig.4.

## 4. Elastic necking with non-monotonous dependence $S(\lambda)$

This Section considers the equilibrium uniaxial stretching a long bar in the form of filament or film, whose dependence $S(\lambda)$ is given by Eq.(15). The analysis shows how in the equilibrium elastic case the present model is incorporated in the general continuum framework established by Ericksen (1975) Coleman (1981,1985) and Coleman and Newman (1988,1990,1992).

The conditions for the $S(\lambda)$ non-monotony, easily established from Eq.(15) read:

$$G_c m \geq G_c c_0 \gg G_a \quad (c_0 \geq 0.3, \ G_c m / G_a \geq 10). \tag{16}$$

Here in the first inequality (16) we used the formula for $m$ ($m = c_0 l_k / h$, with $l_k > h$) in Eq.(4). Under condition (16) the derivative $S'(\lambda)$ in (15) has two roots, $\lambda_*$ and $\lambda^*$ ($\lambda_* < \lambda^*$), the first, $\lambda_*$, corresponding to the maximum of $S(\lambda)$, and the second, $\lambda^*$, to the minimum of $S(\lambda)$. Thus in the intervals $(1, \lambda_*)$ and $(\lambda > \lambda^*)$, the function $S(\lambda)$ monotonically increases and it monotonically decreases in the interval $(\lambda_*, \lambda^*)$.

For the SC polymers with long-range ("flexible") secondary bonds in the crystalline blocks, when $l_k \sim H \sim h$ i.e. $m \approx c_0$, and low degree of crystallinity $c_0$, the value of $\lambda_*$ is relatively large ($\lambda_* \approx 2$), but the interval $(\lambda_*, \lambda^*)$ is very narrow. In this case, one can use the limit $J \to \infty$ ($J \gg I_1$) in Eq. (15) and use the classic expression for the extension force known in the rubber elasticity.

In the more realistic case of short-range ("rigid") secondary bonds in the crystalline blocks when $m > 1$, the value $\lambda_*$ is approaching the unity from above, and $\lambda^* \gg \lambda_*$. In this case, the finite extensibility of polymer chains cannot be ignored.

In order to find the actual constant value of the drawing force $S_0$ during necking, we will first use the Ericksen (1975) approach, which considers the transition zone in necking as a cross-sectional jump, and uses for calculations an elastic potential. In the case of semi-crystalline polymers the true elastic strain energy function in the sense of Eq.(6) does not generally exist, since the actual degree of crystallinity cannot be generally represented as a scalar function of strain tensor. However, in the particular case of simple elongation, such a *pseudo-potential* function $\hat{W}(\lambda)$ always exists and is found from the relation $d\hat{W}/d\lambda = S(\lambda)$ as

$$W(\lambda) = \int_1^\lambda S(\xi)d\xi = W_c(\lambda) + W_a(\lambda), \quad \hat{W}_c(\lambda) = G_c c_0 \left( \frac{1-\lambda^{-m+1}}{m-1} - \frac{1-\lambda^{-m-2}}{m+2} \right). \quad (17)$$

Here $W_a(\lambda)$ is represented in Eq.(11). The Gibbs' (pseudo-) free energy function (per mass unit) $\hat{G}$ is now introduced by the common expression:

$$\hat{G} = \hat{W}(\lambda) - S_0 \cdot (\lambda - 1), \quad (18)$$

where $S_0$ is the actual force. Then the equilibrium condition, $\partial \hat{G}/\partial \lambda \big|_{S_0 = const} = 0$, for any monotonically increasing branch of $S(\lambda)$, yields in accordance with (17): $d\hat{W}/d\lambda = S_0$. If $\lambda_1$ and $\lambda_2$ ($\lambda_1 < \lambda_* < \lambda^* < \lambda_2$) are the stretching ratios achieved during necking at the homogeneously deformed "thick" and "thin" ends of an elongated bar under a constant stretching force $S_m$, then according to the Gibbs' rule applied to the elastic bars by Ericksen (1975),

$$G(\lambda_2) - G(\lambda_1) = \int_{\lambda_1}^{\lambda_2} [S(\lambda) - S_m]d\lambda = 0. \quad (19)$$

Here $S_m (= S_0)$ is the Maxwell stress. The graph in Fig.4 illustrates the Gibbs' rule and Eq.(19).

When the values of $\lambda_1$, $\lambda_2$ and $S_m$ are found, the problem of equilibrium necking is practically solved, because one can use the evident relations,

$$A_k = A_0 / \lambda_k, \quad c_k = c_0 / \lambda_k^m \ (k = 1, 2), \quad V = -U\lambda_1 / (\lambda_2 - \lambda_1). \tag{20}$$

Formulae (20) give the values of cross-sectional areas of the bar in initial state, $A_0$, at the thick end, $A_1$, and at the thin end, $A_2$, as well as the corresponding degrees of crystallinity $c_k$ at the both ends and the relation between the velocity at the thin end $U$ and the neck propagation speed $V$, when the thick end is at rest.

When analyzing the transient neck phenomena, one needs first to use the mass balance averaged over cross-section:

$$\partial_t A + \partial_z (uA) = 0. \tag{21}$$

Here both the velocity $u$ and the cross-section area $A$ depend on time $t$ and the axial space coordinate $z$. Additionally, for the slim bars, one can use the 1D kinematical equation valid for pure simple stretching,

$$\lambda^{-1} d\lambda / dt \approx \lambda^{-1}(\partial_t + u\partial_z) = \dot{\varepsilon} \approx \partial_z u.$$

This equation can be represented in the equivalent form,

$$\partial_t (\lambda^{-1}) + \partial_z (u\lambda^{-1}) = 0. \tag{22}$$

Comparing Eqs.(21) and (22), one can find:

$$\lambda A = const = A_0, \tag{23}$$

where $A_0$ is the cross-sectional area of bar before deformation, when $\lambda = 1$. Eq.(23) has the same form of incompressibility condition as in homogeneous elongation deformation of the bar.

We consider further only quasi-stationary regimes when the entire bar configuration moves in the axial $z$ direction as a stationary wave. Then passing to the moving frame of reference by the transformation, $z \to x = z - Vt$, where $V$ is the propagation speed, one can consider all the phenomena as stationary. In this case, Eqs.(21) and (22) are reduced to:

$$f(u-V) = const, \quad \lambda^{-1} \cdot (u-V) = const. \tag{24}$$

According to Coleman and Newman (1988), the approximate governing equation for propagation of elastic neck in the transition inhomogeneous zone can be presented in the integrable form:

$$S(\lambda) - S_0 = 1/2 (d\gamma/d\lambda)\lambda_x + \gamma(\lambda)\lambda_{xx}. \tag{25}$$

Here the lower index "$x$" denotes the derivative with respect to $x$ variable. The positive function $\gamma(\lambda)$ in Eq.(23) was represented in paper by Coleman and Newman (1988) by derivatives of corresponding 3D elastic potential with respect to the principal values of the Finger strain tensor $\underline{\underline{B}}$. As shown in papers by Coleman (1981,1985) and Coleman and Newman (1988,1990,1992), Eq.(25) is valid with the accuracy of $O(b\lambda_x)^4$, where $b$ is a characteristic thickness of elastic bar (the diameter of filament or thickness or film). Eq.(25) asserts the parabolic character of the governing equation for unsteady necking, the term with the second order space derivative providing the stabilization of neck propagation. An evident variation treatment of the problem has also been considered in papers by Coleman (1981,1985) and Coleman and Newman (1988,1990,1992) 2](see also Leonov (1990)). Multiplying Eq(25) by $\lambda_x$ and integrating it over $x$ yields the first integral of Eq.(25):

$$1/2 \gamma(\lambda)\lambda_x^2 - \hat{G}(\lambda) \equiv 1/2 \gamma(\lambda)\lambda_x^2 - \hat{W}(\lambda) + S_0 \cdot (\lambda - 1) = const. \tag{26}$$

The left-hand side of Eq.(26) represents the density of "Hamiltonian" for the elastic bar.

As mentioned, the 3D elastic potential (strain energy function) does not generally exist in our model. Therefore to calculate the function $\gamma(\lambda)$ as proposed in papers by Coleman (1981,1985) and Coleman and Newman (1988,1990,1992) we use the derivatives of the strain energy function (14) at constant degree of crystallinity. Calculating in this way the function $\gamma(\lambda)$ according to paperby Coleman and Newman (1988) (Eq.(44) there) yields:

$$\gamma(\lambda) = \frac{b^2}{32\lambda^2} \left( \frac{G_c/\lambda^m}{\lambda^2 + \lambda + 1} + \frac{G_a \lambda(J-3)}{J - \lambda^2 - 2\lambda^{-1}} \right). \tag{27}$$

Eq.(27) holds for uni-axially extended samples during necking for both the elastic filaments and films. The boundary conditions for necking phenomena described by Eqs.(22), (26) are:

$$x \to -\infty: \quad u = 0, \quad \lambda_x = 0 \quad (\lambda = const = \lambda_1);$$
$$x \to +\infty: \quad u = const = U, \quad \lambda_x = 0 \quad (\lambda = const = \lambda_2 \ (> \lambda_1));$$
(28)

In this treatment, the parameter $S_0$ is considered as an eigenvalue of the boundary problem (25), (28), and the parameters $\lambda_1$ and $\lambda_2$ are proved to be found to satisfy the Ericksen condition (19), and therefore they are unique functions of $S_0$. The asymptotic treatment of the boundary problem as an "infinite" reflects the exponential decay of the solution at $x \to \pm\infty$. We assume that the thick end of the bar is at rest and the thin one is extended with a constant speed $U$. Using the boundary conditions (28) also evidently yields the formulae (20), where one should additionally employ the incompressibility condition, $\lambda = A_0 / A$. Substituting this condition into Eq.(22) and using the formula (20) for the speed of neck propagation $V$, yields the relation between the local velocity $u$ and stretching ratio $\lambda$ in necking as follows,

$$u(x) = U \frac{\lambda(x) - \lambda_1}{\lambda_2 - \lambda_1}. \tag{29}$$

Eq.(29) being based only on continuum kinematics does not depend on the constitutive relations and holds in both the equilibrium and non-equilibrium cases. It is also evident from Eq.(26) that elastic inhomogeneous necking problem has an analytical solution discussed earlier by Coleman (1981,1985) and Coleman and Newman (1988,1990,1992).

## 5. Forced elasticity and non-equilibrium effects in polymer necking

According to the present model, SC polymers can be viewed as a type of cross-linked elastomers, whose "cross-links" are represented by crystalline blocks unfolded during necking process. These specific cross-links still exist after necking in almost amorphous polymer with oriented parts of macromolecules between the "cross-links". Therefore one can expect that after a rapid unloading, the stretched sample from a SC polymer will almost completely recover its initial length. This scenario does not happen, however, mostly because the amorphous phase in a SC polymer is not in rubbery but rather in glassy state where the mobility of large polymer chains is suppressed by a low temperature. That is why the samples of SC polymers, heated after necking to a (rubbery) temperature below the melting point, demonstrate almost complete recovery (Lazurkin and Fogel'son (1951), Lazurkin (1958), Gent and Geong (1986), Gent and Madan (1989)). This explanation, however, does not answer another question, why the irreversible, viscous, effects are not observed in polymer necking at relatively low speed of extension, although the viscosities in amorphous phase are extremely high? To answer this question we assume (see also Lazurkin and Fogel'son (1951), Lazurkin (1958)) that at high level stresses, observed in polymer necking, the Eyring's activation mechanism (e.g. see Halsey, White and Eyring (1945)) essentially decreases the viscosity in amorphous phase to such a level that dissipative effects and related to them dissipative heat generation are negligible as compared to the leading elastic effects. As soon as the sample is isothermally released from load, the viscosity in amorphous phase, due to a relatively low temperature, jumps back to such a high value that the deformation in the sample could be recovered only in astronomical times.

We now propose a simple viscoelastic model that takes into account both, the elastic and inelastic kinetic effects. We assume that the longitudinal force stretching a SC polymeric bar consists of sum of elastic and inelastic constituents, i.e. $S = S^e + S^i$, where during the neck propagation, when $S = S_0$, $S^e = S(\lambda)$ described by Eq.(15), and $S^i$ is a viscous force,

$$S^i = \eta(T,\sigma)\dot\varepsilon, \quad \eta(T,\sigma) = \eta^* \exp\left(\frac{E - v\sigma}{RT}\right), \quad \sigma = \lambda S_0, \quad \dot\varepsilon = u_x. \qquad (30)$$

Here $\dot\varepsilon$ is the stretching rate, and the viscosity $\eta(T,\sigma)$ is represented in the Eyring (1945) form, where $E$ is the activation energy, $\sigma$ is the actual stress, $T$ is the temperature, $R$ is the gas constant, and the "activated" volume $v$ and pre-exponential factor $\eta^*$ being considered as constant material parameters. Adding the elastic and viscous forces with the use of equations (23), (27) and (28) yields:

$$S_0 = S(\lambda) - 1/2\gamma_\lambda'(\lambda)\lambda_x - \gamma(\lambda)\lambda_{xx} + \lambda_x[\eta_0(T)/\Delta\lambda]U \exp(-v\lambda S_0/RT). \qquad (31)$$

$$\eta_0(T) = \eta^* \exp(E/RT); \quad \Delta\lambda = \lambda_2 - \lambda_1.$$

For steady neck propagation, the boundary conditions to Eq.(31) are the same as shown in Eq.(28). Here parameter $S_0$, the actual force acting on polymeric bar during the steady neck propagation, is treated once again as the eigenvalue of the boundary problem (28),(31). Note that the value of $\Delta\lambda$ is a unique function of the eigenvalue $S_0$. When the difference in the $\lambda$ derivatives of both ascending branches of $S(\lambda)$ is a monotonous function of $\lambda$, the function $\Delta\lambda(S_0)$ is monotonous.

Multiplying Eq.(31) by $\lambda_x$ results in:

$$(1/2)[\gamma(\lambda)\lambda_x^2]_x = [S(\lambda) - S_0]\lambda_x - \lambda_x^2[\eta_0(T)/\Delta\lambda]U \exp(-\nu\lambda S_0/RT). \tag{31a}$$

Therefore on the phase plane, $\lambda_x(\lambda)$, Eq. (31a) takes the form:

$$(1/2)[\gamma(\lambda)\lambda_x^2]_\lambda' = S(\lambda) - S_o - \lambda_x[\eta_0(T)/\Delta\lambda]U \exp(-\nu\lambda S_0/RT). \tag{32}$$

Introducing the new functions,

$$w = \lambda_x\sqrt{\gamma(\lambda)}, \quad a(S_0,\lambda) = \frac{\eta_0 U}{\Delta\lambda\sqrt{\gamma(\lambda)}} \exp(-\nu S_0\lambda/RT), \tag{33}$$

reduces Eq.(32) and corresponding boundary conditions for the case of neck propagation to the form:

$$\frac{dw}{d\lambda} = \frac{S(\lambda) - S_0 + wa(S_0,\lambda)}{w} \quad (\lambda_1 < \lambda < \lambda_2); \quad w(\lambda_1) = w(\lambda_2) = 0. \tag{34}$$

It should be noted that in the most interesting case of small enough viscosity (or large enough $S_0$), function $a(S_0,\lambda)$ introduced in Eq.(33) is a decreasing function of $\lambda$, except, perhaps, a small vicinity of the point $\lambda_1$.

Elementary analysis of phase diagram for Eq.(34) reveals the three singular points, $\lambda_1, \lambda_*$ and $\lambda_2$ ($\lambda_1 < \lambda_* < \lambda_2$). The points $\lambda_1$ and $\lambda_2$ where $S'(\lambda_{1,2}) > 0$ are the saddle points, while the point $\lambda_*$ where $S'(\lambda_*) < 0$, represents either unstable focus when the viscosity is small $\left(4S'(\lambda_*) + a^2(\lambda) < 0\right)$ or unstable knot otherwise. A qualitative picture of the phase trajectories corresponding to the case of necking with small viscosity (the point $\lambda_*$ is unstable focus) is shown in Fig.5. Here the line $S = S_0$ corresponds to $w \equiv 0$, the medium thick solid line shows the plot $S(\lambda)$, the medium thick dashed line depicts the plot $w*(\lambda)$ corresponding to zero of the numerator in Eq.(34), the thin solid lines represent the trajectories with boundary conditions different from that in (34), and the thick solid line depicts the solution of the boundary problem (34). This solution represents a separatrix going out of the saddle point $\lambda_1$ as its unstable mustache and coming into the saddle point $\lambda_2$ as its stable mustache. The existence and uniqueness of solution for the nonlinear boundary value problem (34) with egenvalue $S_0$ has also been proved. The key fact here is the proof that the "length" of the positive part of separatrix, which goes from the first saddle point with $\lambda = \lambda_1$ as the unstable moustache, is a monotonically increasing and continuous function of the parameter $S_0$.

Multiplying Eq.(34) by $w(\lambda)$ and integrating the result from $\lambda_1$ to $\lambda_2$ with the use of boundary condition in (34) yields:

$$w(\lambda) = 2\left(\int_{\lambda_1}^{\lambda}[S(q)-S_0]dq + \int_{\lambda_1}^{\lambda}a(S_0,q)w(q)dq\right)^{1/2}, \qquad (35)$$

Assigning $\lambda = \lambda_2$ results in the integral relation:

$$\int_{\lambda_1}^{\lambda_2}[S(\lambda)-S_0]d\lambda + \int_{\lambda_1}^{\lambda_2}a(S_0,\lambda)w(\lambda)d\lambda = 0, \qquad (36)$$

Since the functions $S(\lambda), a(\lambda)$ and $w(\lambda)$ are positive, Eq.(36) shows that with the viscous term, $S_0$ is larger than the Maxwell stress $S_m$ defined by Eq.(19).

When the viscosity term is so small that it can be treated as a disturbance, the approximate solution is:

$$w(\lambda) \approx 2\left(\int_{\lambda_1}^{\lambda}[S(q)-S_0]dq + \int_{\lambda_1}^{\lambda}a(\hat{S}_0,q)\hat{w}(q)dq\right)^{1/2}; \quad \hat{w}(\lambda) = 2\left(\int_{\hat{\lambda}_1}^{\lambda}[S(q)-\hat{S}_0]dq\right)^{1/2}. \quad (37)$$

Here the overcap symbolizes the inviscid solution of the problem. The first formula in Eq. (37) has sense if the function $\hat{w}(\lambda)$ is defined as taking zero value outside the interval $(\lambda_1,\lambda_2)$. Parameters $\lambda_1, \lambda_2$ and $S_0$ are uniquely defined as:

$$S_0 \approx \int_{\lambda_1}^{\lambda_2}[S(\lambda) + a(S_0,\lambda)\hat{w}(\lambda)]d\lambda. \qquad (38)$$

We finally consider the thermal effects that have been observed by Muller (1949) and Marshall and Thompson (1954) in relatively fast propagating necks. We assume that a noticeable temperature gradient exists only in the short necking region where $\lambda_x$ changes highly, and therefore the adiabatic approach to heat phenomena is appropriate. In this case, the heat equation is written in the form:

$$\rho c_p [u(T - T_1)]_x \approx \lambda_x^2 (U/\Delta\lambda)^2 \eta_0(T) \exp(-\nu S_0 \lambda / RT). \tag{39}$$

Here $T_1$ is the constant temperature of the bar thick end.

It should be noted that in the heat equation for the rubber-like material where the assumption of entropic elasticity is commonly made, the heat source term is not the dissipation but the mechanical power. However in our case, when the amorphous phase in SC polymers is in glassy state, the internal energy could depend not only a temperature but also on strains. Therefore the heat source term in Eq.(39) represents the dissipation in the system.

Adding Eq.(39) multiplied by $\Delta\lambda/U$ to Eq.(31a) results in:

$$1/2[\gamma(\lambda)\lambda_x^2]_x + (\rho c_p \Delta\lambda / U)[u(T - T_1)]_x = [S(\lambda) - S_0]\lambda_x.$$

Integrating this equation with the use of boundary conditions (28) over the entire necking region, $-\infty < x < \infty$, yields:

$$\rho c_p \Delta T = \int_{\lambda_1}^{\lambda_2} [S(\lambda) - S_0] d\lambda / \Delta\lambda. \qquad (\Delta T = T_2 - T_1) \tag{40}$$

Here the common simplifying assumption has been made that $\rho c_p \approx const$. Eq.(40) has the evident physical sense: the total temperature increase in the necking is proportional to the deviation of total change of the Gibbs' potential (18) from its equilibrium (pure elastic) value. Eq.(40) also shows that this temperature change is proportional to the total dissipation in the propagating neck.

## 6. Concluding Remarks

The theoretical approach developed in this paper for semi-crystalline polymers attempted to answer several questions, which remained unanswered during the years.

(i) It was demonstrated that a simple model of unfolding polymer crystals results in non-monotonous behavior of the dependence of stretching force on the stretching ratio, which has been demonstrated as a formal reason for occurrence of necking in many previous papers;

(ii) A simple approach was also developed for modeling viscoelastic, dissipative phenomena in necking which at least qualitatively described the mechanism of forced elasticity;

(iii) Both the above models were included in the well-elaborated general continuum formalism.

Yet at least two important necking problems remain unresolved. The *first one* is related to observed occurrence of two consecutive necks in some SC polymers. This usually happens when the SC polymer has a spherulite structure. Then dismantling the spherulites and converting them in the set of slightly oriented crystalline blocks and amorphous phase would be consider as the reason for occurrence of the first necking (e.g. see Peterlin and Olf (1966), Statton (1967)). Then the second necking will follow the mechanism of unfolding the crystalline blocks developed in this paper. The *second problem* is related to unknown mechanism of very small amplitude striations occurred at the thin end of sample, right behind the neck region (see Fig.3 in paper by Gent and Madan (1989)).

**Acknowledgement**

The author is grateful to Dr. M. Cakmak for fruitful discussions.

**References**

Allison, S.W., Ward, I. M. 1967. Cold drawing of poly(ethylene terephthalate).Br. J. Appl. Phys. 18, 1151-1164.

Antman, S.S. 1973. Non-uniqueness of equilibrium states for bars in tension. J. Math. Anal. 44, 333-349.

Antman, S.S. 1974. Qualitative Theory of Ordinary Differential Equations of Non-Linear

Elasticity, S. Nemat-Vasser Ed., Pergamon, New York, pp.58-101.

Barenblatt, G.I. 1964. On the neck propagation under tension of polymeric samples. Appl. Math. Mech. (Russian), 28 (6), 1048-1060.

Barenblatt, G.I., Entov, V.M., Segalov, A.E. 1969. On thermal mechanism of cold drawing of polymers, Proc. IUTAM Symp. on Thermoelasticity, East Kilbrite, 1968, Springer, Vienna..

Bartenev, G.M. 1964. Determination of the activation energy of polymeric viscous flow from experimental data. Vysokomolekul. Soedin. 6 (2), 335-340.

Bernstein, B., Zapas, L.J. 1981. Stability and cold drawing of viscoelastic bars. J. Rheol. 25, 83-94.

Brauer, P., Muller, F.H. 1954. Temperature elevation in the zone of flow during cold drawing [of polymers]. Koll. Z. 135, 65-67.

Burke, M.A., Nix, W.D. 1979. A numerical study of necking in the plane tension test. Int. J. Solids Struct. 15, 379-393.

Carothers, W.H., Hill, J. W. 1963. Polymerization and ring formation. XV. Artificial fibers from synthetic linear condensation superpolymers**.** J. Am. Chem. Soc. 54, 579-1587.

Coleman, B.D. 1981. Necking and drawing in polymeric fibers under tension. Archive of Rational Mechanics and Analysis. 83, 115-137.

Coleman, B.D. 1984. A Phenomenological theory of the mechanics of cold drawing, in: Orienting Polymers, J.L. Ericksen Ed., Springer, New York. 76.

Coleman, B.D. 1985. On the cold drawing of polymers. Compt. Math. Appl., 11, 35-65.

Coleman,B.D., Newman, D.C. 1988. On the rheology of cold drawing. I. Elastic materials. J. Polym. Sci. B 26, 1801-1822.

Coleman,B.D., Newman, D.C. 1990. Mechanics of neck formation mechanism for cold drawing of elastic films. Polym. Eng. Sci. 30, 1299-1302.

Coleman,B.D., Newman, D.C. 1992. Rheology of neck formation in the cold drawing of polymeric fibers. J. Appl. Polym. Sci. 45, 997-1004.

Coleman,B.D., Newman, D.C. 1992, On the rheology of cold drawing. II. Viscoelastic materials. J. Polym. Sci. Part B 30, 25-47.

Crissman, J.M., Zapas, I.J. 1974. Creep failure and fracture of polyethylene in uniaxial extension. Polym. Eng. Sci. 19, 99-103.

Ericksen, J.L. 1975. Equilibrium of bars. J. Elasticity, 5, 191-201.

Gent, A.N., Jeong, J. 1986. Plastic deformation of crystalline polymers. Polym. Eng. Sci. 26, 285-289.

Gent, A.N., Madan, S. 1989. Plastic yielding of partially crystalline polymers. J. Polym. Sci., B:Polym Phys. 27, 1529-1542.

Gent, A.N. 1996. A new constitutive relation for rubber. Rubber Chem. Technol., 69, 59-61.

Gul',V.Ye, Kuleznyov, V.N. 1972. The Structure and Mechanical Properties of Polymers, 2$^{nd}$ ed. (Russian), High Education, Moscow.

Halsey, G., White, H.J., Eyring, H. 1945. Mechanical properties of textiles. I. Text. Res. J., 15, 295-311.

Juska, T., Harrison, I.R. 1982. A proposed plastic deformation mechanism for semicrystalline polymers. Polym. Eng. Rev. 2, 13-28.

Juska, T., Harrison, I.R. 1982a. A criterion for craze formation. Polym. Eng. Sci. 22, 766-776.

Kargin, V.A., Slonimsky, G.L. 1967. Short Course of the Physical Chemistry of Polymers. Khimiya, Moscow.

Kozlov, P.V., Kabanov,V.A., Frolova,A.A. 1959. Certain correlations in the development of uniaxial deformation in crystalline and vitreous films from poly(ethylene terephthalate). M. V. Lomonosov State Univ., Moscow, Vysokomolekulyarnye Soedineniya Vsesoyuz. Khim. Obshchestvo im. D. I. Mendeleeva 1(2), 324-329.

Larson, R.G. 1988. Constitutive Equations for Polymer Melts and Solutions, Butterworth, Boston. Section 5.8.3.

Lazurkin, Yu.S., Fogel'son, R.L. 1951 Nature of the large deformations of high-molecular compounds in the vitreous state. Inst. Phys. Problems, Acad. Sci. U.S.S.R., Moscow, Zhur. Tekh. Fiz. 21, 267-286.

Lazurkin, Yu.S. 1958. Cold-drawing of gass-like and crystalline polymers. J. Polym. Sci.,

30, 595-604.

Leonov, A.I. 1990. The effect of surface tension on stretching of very thin highly elastic filaments. J. Rheol. 34,155-167.

Marshall, I., Thompson, A.B. 1954. The cold drawing of high polymers. Proc. Roy. Soc. London, A 221, 541-557.

Muller, F.H. 1949. Problem of cold-drawing of high-polymeric substances. Koll. Z. 114, 59-61.

Nadai, A. 1950. Theory of Flow and Fracture of Solids, McGraw-Hill, New York.

Needleman, A. 1972. A numerical study of necking in circular cylindrical bars. J. Mech. Phys. Solids, 20 , 111-127.

Orowan, E. 1949. Fracture and strength of solids [metals]. Rep. Progr. Phys.12, 186-232.

Owen, N., 1987. Existence and stability of necking deformations for nonlinearly elastic rods. Archive of Rational Mechanics and Analysis, 98, 357-383.

Peterlin, A., Olf, H.G. 1966. NMR observations of drawn polymers. V. Sorption into drawn and undrawn polyethylene. J. Polym. Sci. A-2, 4, 587-598.

Silling, S.A. 1988. Two-dimensional effects in the necking of elastic bars. J. App. Mech.-Trans. ASME 55, 530-535.

Statton, W.O. 1967. Coherence and deformation of lamellar crystals after annealing. J. Appl. Phys. 38, 4149-4151.

Tager, A.A. 1978. Physical Chemistry of Polymers. Mir, Moscow, USSR.

Truesdell, C., Noll, W. 1992. The Non-Linear Field Theories of Mechanics, 2$^{nd}$ Ed., Springer, New York, Sections 86, 95.

Vincent, P.I. 1960. Necking and cold-drawing of rigid plastics. Polymer. 1, 7-19.

Ward, I.M. 1982. Mechanical Properties of Solid Polymers, 2$^{nd}$ ed., Wiley, New York.

Warner, H.R. Jr. 1972. Kinetic theory and rheology of dilute suspensions of finitely extendible dumbbells. Ind. Eng. Chem. Fundamentals, 11, 379-387.

Whitney, W., Andrews, R.D. 1967. Yielding of glassy polymers: volume effects. J. Polym. Sci. C16,

Zapas, I.J., Crissman, J.M. 1974. An instability leading to failure of polyethylene in uniaxial creep. Polym. Eng. Sci. 19,104-107.

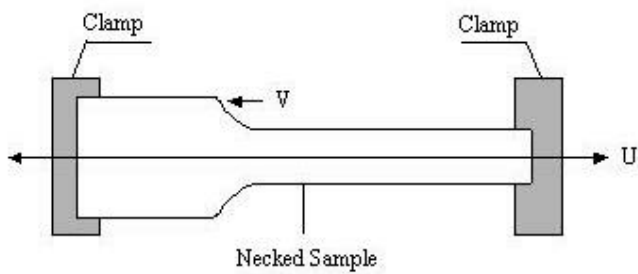

Fig. 1 Sketch of Necking Polymer Sample

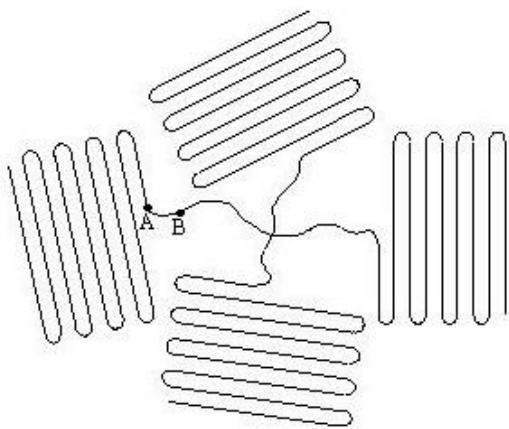

Fig.2 Sketch of Amorphous Crystalline Regions in Semi-Crystalline Polymers
A-B: indicates the macromolecular segment in transition from crystalline to amorphous region

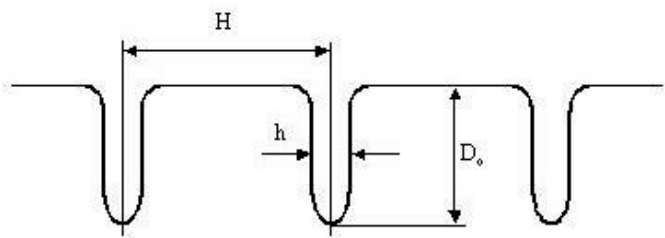

Fig.3 Sketch of Potential Well in a Crystalline Block

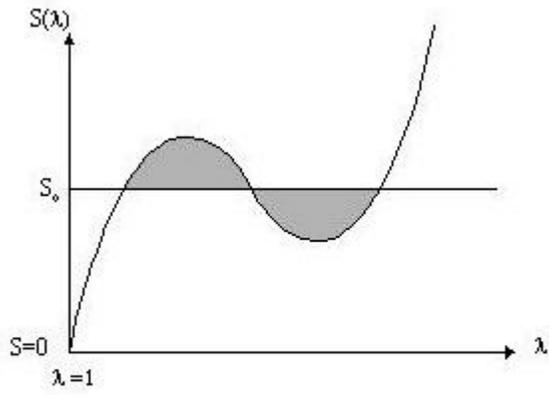

Fig.4 Schematics of Maxwell Stress $S_o$

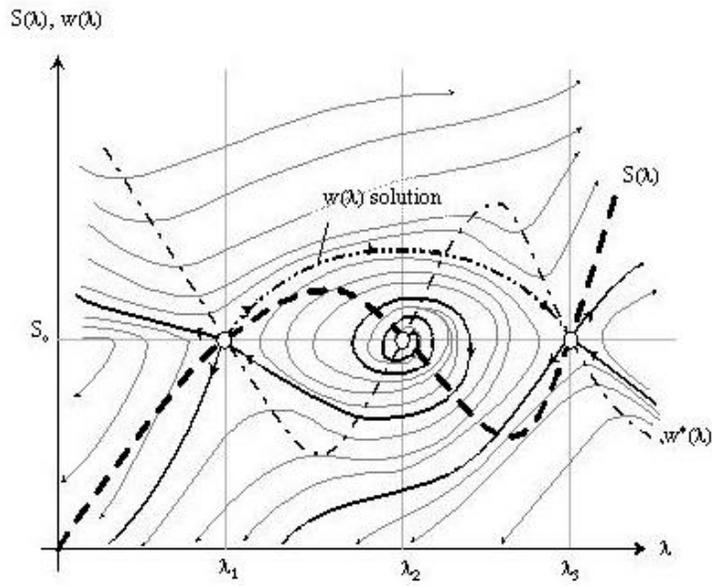

Fig.5 Phase Diagram for Eq.(34)